\documentclass[11pt,a4paper]{article}
\usepackage{subeqnarray}
\usepackage[latin1]{inputenc}
\usepackage[T1]{fontenc}
\usepackage{natbib}

\bibpunct{[}{]}{;}{a}{}{;}

\usepackage[dvips]{graphicx}
\usepackage{rotating}
\usepackage{pstricks}
\makeatletter
\def\fnum@figure{\footnotesize\bfseries%
                 \figurename~\thefigure%
                 \mdseries}
\renewenvironment{figure}
               {\@float{figure}}
               {\end@float}
\def\fnum@table{\footnotesize\bfseries%
                 \tablename~\thetable%
                 \mdseries}
\renewenvironment{table}
               {\@float{table}}
               {\end@float}
\makeatother

\makeatletter
\def\re{\mathop{\operator@font Re}}
\def\im{\mathop{\operator@font Im}}
\makeatother
\newcommand{\I}{\mathrm{i\,}}
\newcommand{\cc}{\overline}
\newcommand{\vp}{{\bf U}}
\newcommand{\vpa}{\vp_+}
\newcommand{\vpb}{\cc{\vp}_+}
\newcommand{\chem}[1]{\ensuremath{\mathrm{#1}}}
\newcommand{\stat}[1]{{#1_{\rm s}}}
\newcommand{\negs}{\!\!\!\!\!}
\newcommand{\IIten}{\!\!\!\; : \;\!\!\!}
\newcommand{\IIIten}{\:\!\vdots\!\:}

\newcommand{\sect}{Sec.~}

\newcommand{\eqn}{Eq.~}
\newcommand{\eqns}{Eqs.~}
\newcommand{\fig}{Fig.~}
\newcommand{\figs}{Figs.~}
\newcommand{\tabl}{table~}

\newcommand{\lra}       {\longrightarrow}
\newcommand{\cha}[1]    {\stackrel{#1}{\lra}}
\newcommand{\hp}        {\chem{H^+}}
\newcommand{\bromat}    {\chem{BrO_3^-}}
\newcommand{\bromid}    {\chem{Br^-}}
\newcommand{\hbro}      {\chem{HBrO}}
\newcommand{\cef}       {\chem{Ce^{4+}}}

\newcommand{\cet}       {\chem{Ce^{3+}}}
\newcommand{\brsyr}     {\chem{HBrO_2}}
\newcommand{\molar}     {\scalebox{0.833}{\rm M}}

\begin{document}
\pagestyle{empty}

%%%%%%%%%%%%%%%%%%%%%%%%%%%%%%%%%%%%%%%%%%%%%%%%%%%%%%%
%% Make front page
%%%%%%%%%%%%%%%%%%%%%%%%%%%%%%%%%%%%%%%%%%%%%%%%%%%%%%%
\
\vspace{1.5cm}

\begin{center}
  \LARGE
  Amplitude Equations and

  \vspace{0.5cm}
  Chemical Reaction-Diffusion Systems
  \normalsize
\end{center}

\begin{center}
  \Large
  M.\ Ipsen \hspace{5.0mm} F.\ Hynne \hspace{5.0mm} P.\ G.\ Sørensen

  \vspace{0.4cm}
  \today
  \normalsize
\end{center}

\vspace{0.3cm}
\begin{center}
  \emph{%
    Department of Chemistry, 
    University of Copenhagen,
    H.C.Ørsted Institutet,
    Universitetsparken 5,
    2100 DK-Copenhagen,
    Denmark.
    }
\end{center}

%%%%%%%%%%%%%%%%%%%%%%%%%%%%%%%%%%%%%%%%%%%%%%%%%%%%%%%
%% Abstract
%%%%%%%%%%%%%%%%%%%%%%%%%%%%%%%%%%%%%%%%%%%%%%%%%%%%%%%
\addtocounter{page}{-1}
\begin{abstract}
  The paper discusses the use of amplitude equations to
  describe the spa\-tio-tempo\-ral dynamics of a chemical
  reaction-diffu\-sion system based on an Oregonator model of
  the Belousov-Zhabotinsky reaction. Sufficiently close to a
  supercritical Hopf bifurcation the reaction-diffusion
  equation can be approximated by a complex Ginzburg-Landau
  equation with parameters determined by the original equation
  at the point of operation considered. We illustrate the
  validity of this reduction by comparing numerical spiral
  wave solutions to the Oregonator reaction-diffusion equation
  with the corresponding solutions to the complex
  Ginzburg-Landau equation at finite distances from the
  bifurcation point. We also compare the solutions at a
  bifurcation point where the systems develop spatio-temporal
  chaos. We show that the complex Ginzburg-Landau equation
  represents the dynamical behavior of the reaction-diffusion
  equation remarkably well sufficiently far from the
  bifurcation point for experimental applications to be
  feasible.
\end{abstract}

\vspace*{\fill}
To appear in \emph{Int.\ J.\ Bifurcation and Chaos},
\textbf{7}(9), 1997.

%%%%%%%%%%%%%%%%%%%%%%%%%%%%%%%%%%%%%%%%%%%%%%%%%%%%%%%
%% Main text
%%%%%%%%%%%%%%%%%%%%%%%%%%%%%%%%%%%%%%%%%%%%%%%%%%%%%%%
\newpage
\pagestyle{plain}
\section{Introduction}
During the last two decades, the study of spiral waves and
spatio-temporal chaos in physical, chemical, and biological
systems has received much attention. In physics, spiral
patterns have been studied extensively in hydrodynamic
systems~\citep{CrossHohen}. Recently, spiral formation has
also been studied in liquid
crystals~\citep{LiqCryst}. Biological examples include the
growth of the slime mold \emph{Dictyostelium
discoideum}~\citep{Dicty} and potential differences occuring
on the surface of rabbit heart muscles~\citep{Rabbit}. In
chemical systems, spiral formation and related phenomena like
target wave propagation often occur in oscillatory chemical
reactions which take place in unstirred media (\emph{e.g.} a
petri dish). In this case the spirals appear as concentration
variations over the spatial domain of the system. A well-known
example of a chemical reaction exhibiting such phenomena, is
the Belousov-Zhabotinsky reaction, whose properties in
spatially distributed media has been discussed extensively in
the chemical literature. For further reference, see
Zhabotinsky~[\citeyear{ZhabOverview}] and references
therein. Recently, observations of spiral wave development and
spatio-temporally chaotic patterns close to a supercritical
Hopf bifurcation was reported by Ouyang \&
Flesselles~[\citeyear{CgleNature}].

\vspace{\parsep}%
The state space of chemical systems exhibiting wave and
spatio-temporally chaotic phenomena almost always have a
fairly high dimension. Consequently, realistic modeling of
such phenomena by direct integration of the reaction-diffusion
equation is difficult or sometimes even impossible. For this
reason most works in this field have used simplified
two-dimensional models, often describing excitable media.

\vspace{\parsep}%
For oscillatory systems, the motion in the concentration space
largely takes place on a two-dimensional manifold. In
particular, this condition applies near a supercritical Hopf
bifurcation where, furthermore, the motion in the plane of
oscillations can be described analytically to a very good
approximation. Integrating the reaction-diffusion system close
to a supercritical Hopf bifurcation can be a very time
consuming and difficult task even on a large computer. For
this reason, it is of great importance to introduce techniques
that utilize the characteristics of the Hopf bifurcation in
order to simplify the reaction-diffusion equation (RDE)
without changing the essential properties of the solution. As
shown by Kuramoto~[\citeyear{Kuramoto}], such a description is
offered by the complex Ginzburg-Landau equation (CGLE) which
provides a systematic description of the chemical waves near a
Hopf bifurcation (of the corresponding homogeneous system). He
applied it to the Brusselator, a two-dimensional model for an
abstract chemical system.

\vspace{\parsep}%
The CGLE is an amplitude equation which describes the waves of
the reaction-diffusion system in terms of a complex amplitude.
It offers a number of advantages in addition to a reduction of
the effective dimension of the original reaction-diffusion
equation. It provides an intuitively natural description of
oscillations which facilitates interpretations of results and,
furthermore, introduces a scaling which makes it universal,
depending only on two complex parameters but not \emph{e.g.}
on the distance from the bifurcation point. The scaling also
solves a serious problem associated with the integration of
reaction-diffusion equations, due to the critical slowing down
of the motion in the concentration space which occurs near the
bifurcation point. The description becomes increasingly
precise as the Hopf bifurcation point is approached and is
``exact'' in the limit.

\vspace{\parsep}%
As one moves further away from the bifurcation point the CGLE
description gradually becomes less precise. Hence we wish to
discuss how well the CGLE description approximates realistic
reaction-diffusion systems representing actual chemical
systems at finite distances from a Hopf bifurcation. This
point is very important from an experimental point of view,
since experimental observations of waves and spatio-temporal
chaos require sufficiently large amplitudes and rates of
development for spatial patterns.

\vspace{\parsep}%
The behavior of amplitude equations that apply close to a Hopf
bifurcation have for example been studied by \cite{Huber92},
\cite{Ar93}, and \cite{Nic95}. However, to our knowledge,
there have been no attempts to test how well the CGLE actually
does approximate the behavior of a realistic
reac\-tion-diffusion system close to a supercritical Hopf
bifurcation. By comparing results from simulations of a
specific reaction-diffusion system with those of the
associated CGLE, we show in this paper that the CGLE indeed
gives a very satisfactory description of the original
system. We focus attention on two specific conditions for an
Oregonator model, where spiral patterns and chaotic patterns
can occur when the corresponding homogeneous systems exhibits
simple oscillations with small amplitude.

\section{Review of Theory}
We consider a chemical reaction with $n$ chemical components
taking place under heterogeneous conditions described by the
reaction-diffusion equation. The reaction-diffusion equation
is defined by the rate expression and the diffusion matrix
associated with the chemical system under consideration. Let
the vector ${\bf c}$ describe the concentrations of the $n$
dynamical species that participate in the reaction-diffusion
system. The associated reaction-diffusion equation can then be
written as
\begin{equation}
  \dot{{\bf c}} = {\bf f}({\bf c};\mu) +
                     {\bf D} \!\cdot\! \nabla^2_{\bf r}{\bf c},
  \label{eq:ReacDiff}
\end{equation} 
where ${\bf f}$ is the rate expression and ${\bf D}$ is the
diffusion matrix. Here $\mu$ designates some controllable
parameter which can be varied experimentally and serves as a
bifurcation parameter. Assume that this system has a
homogeneous and stationary solution ${\bf c}_s$.  The dynamics
of small perturbations ${\bf u}(t,{\bf r};\mu) = {\bf
c}(t,{\bf r};\mu) - \stat{\bf c}(t;\mu)$ of the stationary
point can then be described through a Taylor expansion of the
right-hand side of \eqn(\ref{eq:ReacDiff}) by
\begin{equation}
  \label{eq:TaylorExp}
  \dot{\bf u} = {\bf J} \!\cdot\! {\bf u} + \frac{1}{2!}{\bf M\IIten  uu} 
                         + \frac{1}{3!}{\bf N\IIIten uuu} + \ldots + 
  {\bf D} \!\cdot\! \nabla^2_{\bf r}{\bf u}.
\end{equation}
The matrix $J_{ij}=\frac{\partial f_i}{\partial c_j}$ is the
Jacobian matrix, whereas ${\bf M\IIten uu}$ and ${\bf N\IIIten
uuu}$ denote quadratic and cubic forms in ${\bf u}$,
respectively defined by

\begin{subeqnarray}
  ({\bf M\IIten uu})_i  & = & 
  \left.
    \sum_{j,k=1}^{n}
    \frac{\partial^2\!f_i}{\partial c_j\partial c_k}
  \right|_{\stat{\bf c}} \negs u_ju_k,\\
  ({\bf N\IIIten uuu})_i & = & 
  \left.
    \sum_{j,k,l=1}^{n}
    \frac{\partial^3\!f_i}{\partial c_j\partial c_k\partial c_l}
  \right|_{\stat{\bf c}}  \negs u_ju_ku_l.
\end{subeqnarray}

${\bf M}$ is often referred to as the Hessian. The stability
of the stationary state $\stat{\bf c}$ to space-independent
perturbations is determined by the eigenvalues of the Jacobian
matrix associated with the linearization of the rate
expression ${\bf f}({\bf c};\mu)$ in \eqn(\ref{eq:ReacDiff})
around ${\bf c}_s$. In general, a bifurcation or loss of
stability will occur if the real part of one or more of these
eigenvalues change sign from negative to positive. Let us
assume that $\stat{\bf c}$ looses stability via a
supercritical Hopf bifurcation at $\mu=\mu_0$ due to a change
of sign in the real part of a complex eigenvalue
$\lambda=\sigma+\I\omega$ and its complex conjugate. At the
bifurcation point, the eigenvalue $\lambda$ will therefore be
purely imaginary, $\I\omega_0$. In the homogeneous system,
this bifurcation gives rise to small sinusoidal oscillations
of the concentrations. In the limit $\mu \rightarrow \mu_0$,
the frequency of the oscillations will approach the value
$\omega_0$ corresponding to the imaginary part of $\lambda$ at
the bifurcation point.

\vspace{\parsep}%
The equations related to the Hopf bifurcation can be expanded
in a formal parameter $\epsilon$ which often is related to the
amplitude of the oscillations. The quantities ${\bf u}$, ${\bf
J(\mu)}$, ${\bf M(\mu)}$, ${\bf N(\mu)}$, $\lambda(\mu) =
\sigma(\mu) + \I\omega(\mu)$ and the original parameter $\mu$
can be expanded as follows
\begin{equation}
  \label{eq:EpsExpand}
  \begin{array}{rcl}
    & {\bf u} = \epsilon {\bf u}_1 + \epsilon^2 {\bf u}_2
                                   + \epsilon^3 {\bf u}_3 \ldots,
      \;\;\;\;\;\;\,
      {\bf J} = {\bf J}_0 + \epsilon^2 {\bf J}_2
                          + \epsilon^4 {\bf J}_4 \ldots, &\\
    & {\bf M} = {\bf M}_0 + \epsilon^2 {\bf M}_2
                          + \epsilon^4 {\bf M}_4 \ldots,
      \;\;\;\;
      {\bf N} = {\bf N}_0 + \epsilon^2 {\bf N}_2
                          + \epsilon^4 {\bf N}_4 \ldots, &\\
    & \;
      \lambda = \lambda_0 + \epsilon^2 \lambda_2
                          + \epsilon^4 \lambda_4 \ldots, &\\
    & \mu     = \mu_0     + \epsilon^2 \mu_2
                          + \epsilon^4 \mu_4 \ldots &\\
  \end{array}
\end{equation}
where $\lambda_j = \sigma_j + \I\omega_j, j =
0,2,4,\ldots$. The parameter $\epsilon$ can be defined in many
ways. In experimental situations where the bifurcation
parameter is a real physical quantity, it is often convenient
to choose $\epsilon$ in a dimensionless form and eliminate it
from all expressions that are used to interpret
results~\citep{Kosek}.

\vspace{\parsep}%
Consider the behavior of the system in states that locally are
close to the homogeneous oscillatory solution. One can
prove~\citep{Kuramoto} that close to the Hopf bifurcation
point the dynamics of the $n$ chemical species ${\bf c}(t,{\bf
r};\mu)$ can be approximated by the following equation
\begin{equation}
  \label{eq:ConcDyna}
  {\bf c}(t,{\bf r};\mu) = \stat{\bf c} + \epsilon
  \left(
    W(\tau,{\bf s})e^{\I\omega_0t}\vpa + 
    \cc{W}(\tau,{\bf s})e^{-\I\omega_0t}\vpb
  \right).
\end{equation}
Here $\vpa$ is the complex right eigenvector associated with
the bifurcating eigenvalue $\I\omega_0$ at the bifurcation
point $\epsilon = 0$ ($\mu=\mu_0$). The complex amplitude
$W(\tau,{\bf s})$ of \eqn(\ref{eq:ConcDyna}) must satisfy the
complex Ginzburg-Landau equation
\begin{equation}
  \label{eq:GinzLand}
  \frac{\partial W}{\partial \tau} =
  \lambda_2 W - g |W|^2W + d \nabla^2_{\bf s}W,
\end{equation}
which depends on $t$, ${\bf r}$, and $\epsilon$ through the
time and space variables $\tau$ and ${\bf s}$, scaled with
$\epsilon$ as $\tau = \epsilon^2 t$ and ${\bf s} = \epsilon\,
{\bf r}$. One purpose of the scaling is to obtain a
description that is independent of the distance from the
bifurcation point (as long as the approximation is
applicable). The use of the scaled time $\tau$ solves the
problem of critical slowing down exhibited when the system
approaches the bifurcation point. In other words, the closer
we get to the bifurcation point, the longer the transient time
for pattern development. As we shall see, this fact can indeed
cause difficulties in numerical studies of the original
reaction-diffusion equation, which are absent when the CGLE is
used.

\vspace{\parsep}%
The complex parameter $d$ in the CGLE is determined by the
equation
\begin{equation}
  \label{eq:Solv1}
  d = \vpa^\ast \!\cdot\! {\bf D} \!\cdot\! \vpa,
\end{equation}
where $\vpa^\ast$ is the left eigenvector corresponding to the
bifurcating eigenvalue $\I\omega_0$ at the bifurcation
point. The left eigenvector $\vpa^\ast$ has been normalized to
fulfil the relation $\vpa^\ast \!\cdot\! \vpa=1$. To find the
complex coefficient $g$ one first determines two vectors ${\bf
F}_{20}$ and ${\bf F}_{11}$ by solving the two linear
equations~\citep{Kuramoto}
\begin{subeqnarray}
  \label{eq:SolvPrel}
  ({\bf J}_0 - 2\I\omega_0{\bf I}) \!\cdot\! {\bf F}_{20} & = & 
  -\frac{1}{2}{\bf M}_0\IIten\vpa\vpa,\\
  {\bf J}_0 \!\cdot\! {\bf F}_{11} & = & -{\bf M}_0\IIten\vpa\vpb.
\end{subeqnarray}
The complex parameter $g$ can be found by the
expression~\citep{Kuramoto}
\begin{equation}
  \label{eq:SolvCond}
  g = -\vpa^\ast \!\cdot\! {\bf M}_0\IIten\vpa{\bf F}_{11} -
  \vpa^\ast \!\cdot\! {\bf M}_0\IIten\vpb{\bf F}_{20} -
  \frac{1}{2}\vpa^\ast \!\cdot\! {\bf N}_0\IIIten\vpa\vpa\vpb.
\end{equation}
In \eqn(\ref{eq:SolvPrel}) and \eqn(\ref{eq:SolvCond}), the
tensors ${\bf J}_0$, ${\bf M}_0$, and ${\bf N}_0$ denote the
values of ${\bf J}$, ${\bf M}$, and ${\bf N}$ at the
bifurcation point $\mu=\mu_0$.

\vspace{\parsep}%
The complex Ginzburg-Landau equation is often presented in the
form
\begin{equation}
  \label{eq:TransCGLE}
  \frac{\partial W}{\partial \tau} =
  W - (1+\I\beta)|W|^2W + (1+\I\alpha)\nabla^2_{\bf s}W.
\end{equation}
Here $\alpha = \frac{d''}{d'}$ and $\beta = \frac{g''}{g'}$,
where $d'=\re d$, $d''=\im d$, $g'=\re g$, and $g''=\im
g$. One obtains this dimensionless version of the complex
Ginzburg-Landau equation by performing the variable change
$\tau \rightarrow \sigma_2^{-1}\tau$, ${\bf s} \rightarrow
\sqrt{\frac{d'}{\sigma_2}}{\bf s}$, $W \rightarrow
\sqrt{\frac{\sigma_2}{|g'|}}\exp(\I\frac{\omega_2}{\sigma_2}
\tau)$.

\vspace{\parsep}%
One can easily show that in a one-dimensional spatial domain
of infinite length \eqn(\ref{eq:TransCGLE}) admits of plane
wave solutions of the form
\begin{equation}
  \label{eq:PlaneWAV}
  W_Q(x,t) = R_Q e^{\I(Qx-\omega_Qt)},
\end{equation}
where the frequency $\omega_Q$ and amplitude $R_Q$ have to
satisfy the relations
\begin{subeqnarray}
  \label{eq:DispRel}
       R_Q & = & \sqrt{1-Q^2},\;\;\; |Q| < 1,\\
  \omega_Q & = & \beta Q^2 + \alpha (1-Q^2).
\end{subeqnarray}
By considering perturbations of the plane wave solutions, one
can prove that these become unstable when $\alpha$ and $\beta$
satisfy the relation \mbox{$1+\alpha\beta<0$}. The quantity
\mbox{$1+\alpha\beta=0$} determines the boundary of the
Benjamin-Feir instability which is associated with
spatio-temporally chaotic behavior in the system. Since spiral
wave solutions far from the spiral core asymptotically
approach a plane wave locally, one would expect that crossing
the boundary of the Benjamin-Feir instability also implies a
breakdown of spiral waves and perhaps development of chaotic
behavior. In \sect\ref{sec:numerical} we shall present
examples showing the breakdown of spiral wave solutions in
both the reaction-diffusion equation and the Ginzburg-Landau
equation.

\section{Model description}
As a model for a real chemical system we have chosen the three
dimensional Oregonator~ \citep{OregOrig}. The Oregonator is
based on the so-called FKN-mechanism, which provided the first
successful explanation of the chemical oscillations that occur
in the BZ-reaction~\citep{FKNorig}. During the last two
decades, the Oregonator model has been modified in many ways
by inclusion of additional chemical reaction steps or by
changing the rate constants. The model used in this study
consists of the original Oregonator mechanism
\begin{subeqnarray}
  \label{eq:OrgReac}
  \bromat + \bromid + 2\hp & \cha{k_1} & \brsyr + \hbro\\
  \brsyr  + \bromid        & \cha{k_2} & 2\hbro\\
  2\brsyr                  & \cha{k_3} & \brsyr + \bromat + \hp\\
  \bromat + \brsyr + \hp   & \cha{k_4} & 2\brsyr + 2\cef\\ 
  \cef                     & \cha{k_5} & f \bromid + \cet
\end{subeqnarray}
with rate constants $k_1,\ldots,k_5$ from Field \&
F\"{o}rsterling~[\citeyear{FFrate}]. The quantity $f$ is a
stoichiometric factor. For notational simplicity we introduce
$A = [\bromat]$, $H = [\hp]$, $X = [\brsyr]$, $Y = [\bromid]$,
and $Z = [\cef]$, where [S] is the concentration of species
S. For relevant experimental conditions we may assume that
[$\bromat$] and [$\hp$] are constant. By applying mass law
kinetics, the time and space variation of the spatially
distributed Oregonator model can be described by the following
three coupled partial differential equations

\begin{subeqnarray}
  \label{eq:OrgDiff}
  \frac{\partial X}{\partial t} & = & 
  k_1AH^2Y - k_2HXY - 2k_3X^2 + k_4AHX + D_X\nabla_{\bf r}^2X, \\
                                &   &   \nonumber\\
  \frac{\partial Y}{\partial t} & = & 
  -k_1AH^2Y - k_2HXY + k_5fZ + D_Y\nabla_{\bf r}^2Y,          \\
                                &   &   \nonumber\\
  \frac{\partial Z}{\partial t} & = &  
  2k_4AHX - k_5Z + D_Z\nabla_{\bf r}^2Z,
\end{subeqnarray}

where $D_X$, $D_Y$, and $D_Z$ are the diffusion constants of
the species $\brsyr$, $\bromid$, and $\cef$ respectively (for
dilute solutions, the diffusion matrix is diagonal to a good
approximation). For a thorough discussion of the chemistry on
which the Oregonator is based, the reader is referred to
Tyson~[\citeyear{Tyson}]. The values of the stoichiometric
factor $f$ \citep{HopfQuench}, of the rate constants and of
the diffusion constants used in the numerical studies are
shown in \tabl\ref{tab:RateVals}. For the diffusion constants,
we have used estimates by Hynne \&
Sørensen~[\citeyear{PgsCgle}].

\section{Numerical simulations}
\label{sec:numerical}
The parameters $\alpha$ and $\beta$ in
\eqn(\ref{eq:TransCGLE}) have a complicated functional
dependence on the parameters that characterize the Hopf
bifurcation in the original reaction-diffusion system as is
evident from \eqns(\ref{eq:Solv1})--(\ref{eq:SolvCond}). If a
parameter in the original model system is changed slightly, a
new Hopf bifurcation can be found by adjusting another
parameter, corresponding to the fact that the Hopf bifurcation
is a generic co-dimension one bifurcation. The Ginzburg-Landau
parameters $\alpha$ and $\beta$ will also change due to
\eqns(\ref{eq:Solv1})~and~(\ref{eq:SolvCond}). It is therefore
convenient to study the variation of the parameters $\alpha$
and $\beta$ along a branch in a Hopf bifurcation diagram where
the Hopf bifurcation is continued as a function of two
appropriate parameters in the particular system. For the
present Oregonator model, we have chosen the concentrations of
$\bromat$ and $\hp$ as bifurcation parameters, since these
easily can be varied experimentally and are almost constant
during an experiment. The bifurcation diagram obtained for the
Oregonator model is shown in \fig\ref{fig:BifDiagram}. The
curve consists of two branches representing super- and
subcritical Hopf bifurcations shown as the solid and dashed
parts of the curve in \fig\ref{fig:BifDiagram}. This change of
stability occurs at \chem{[\bromat] = 0.269\;\molar} and
\chem{[\hp] = 0.176\;\molar} (\chem{\molar = mol\;dm^{-3}}).

\vspace{\parsep}%
Since \eqn(\ref{eq:GinzLand}) applies to a supercritical Hopf
bifurcation only, our attention will be limited to the
supercritical branch in the bifurcation diagram. One can
calculate the parameters $\alpha$ and $\beta$ from
\eqns(\ref{eq:Solv1})~and~(\ref{eq:SolvCond}) along the
bifurcation curve plotted as functions of $[\bromat]$ in
\fig\ref{fig:AlphaPlot}. We see that the diffusion parameter
$\alpha$ changes slightly as the Hopf bifurcation point varies
with the bifurcation parameters $[\bromat]$ and $[\hp]$. This
variation is due to the fact that the directions of the right
and left eigenvectors of the Jacobian matrix ${\bf J}_0$
change with the Hopf bifurcation point. The coefficient
$\beta$ shows a more dramatic change along the bifurcation
curve. At the right-hand edge of the supercritical branch,
$\beta$ diverges to infinity because the denominator of
$\beta=\frac{g''}{g'}$ becomes zero at the transition from
supercritical to subcritical behavior. An extended
Ginzburg-Landau approach may still apply close to the
singularity if a quintic term is added to the complex
Ginzburg-Landau equation. However, this modification will not
be considered in this paper.

\vspace{\parsep}%
A crucial term which can be calculated from the values of
$\alpha$ and $\beta$ along the Hopf bifurcation branch is the
Benjamin-Feir parameter \mbox{$1+\alpha\beta$}, which
determines whether plane waves are unstable
(\mbox{$1+\alpha\beta<0$}) to small perturbations in one space
dimension. The variation of \mbox{$1+\alpha\beta$} along the
supercritical part of the Hopf bifurcation curve is exhibited
in \fig\ref{fig:BFPlot} which shows that a Benjamin-Feir
instability starts at \chem{[\bromat]_{BF} = 0.073\;\molar}
and \chem{[\hp]_{BF} = 0.348\;\molar}. For \chem{[\bromat] <
[\bromat]_{BF}} and \chem{[\hp] > [\hp]_{BF}}, we may
therefore expect to find stable spiral wave solutions to the
reaction-diffusion system as well as to the complex
Ginzburg-Landau equation. Spiral waves are expected to become
unstable when \chem{[\bromat] > [\bromat]_{BF}} and
\chem{[\hp] < [\hp]_{BF}}. (Recall that \chem{[\bromat]} and
\chem{[\hp]} vary together on the Hopf bifurcation curve.)

\vspace{\parsep}%
We have selected two points on the (supercritical) Hopf
bifurcation curve for comparisons of the (Oregonator based)
reaction-diffusion system with the corresponding complex
Ginzburg-Landau equation as indicated on
\fig\ref{fig:BifDiagram}. The characteristic Ginzburg-Landau
parameters for the two points are shown in
\tabl\ref{tab:BifVals}. The points differ in the sign of the
Benjamin-Feir parameter, \mbox{$1+\alpha\beta$}; they are
marked on \fig\ref{fig:BFPlot}. The complex Ginzburg-Landau
equation applies near a Hopf bifurcation and is scaled so that
its form is independent of the distance from the bifurcation
point. Thus, $\alpha$ and $\beta$ are defined at the
bifurcation. However, we want to compare solutions to the
complex Ginzburg-Landau equation which is independent of the
distance from the bifurcation point with those of the
reaction-diffusion system for which the distance from the
bifurcation point is important. In the latter case, one must
work at definite (finite) distances from the bifurcation
curve. The points are chosen with the following considerations
in mind: If a point is too close to the bifurcation curve,
significant spatial changes in amplitude take so long
(compared to a local period of oscillation) that the numeric
solution of the reaction-diffusion equation becomes
practically impossible. On the other hand, the points should
be close enough to the bifurcation for the complex
Ginzburg-Landau equation to apply to a reasonably good
approximation.

\vspace{\parsep}%
One way, by which one can estimate a reasonable choice for the
distance from the bifurcation, is to study how well the
Stuart-Landau equation, approximates the true limit cycle
solution for the homogeneous chemical reaction system. The
Stuart-Landau equation is simply the complex Ginzburg-Landau
equation without the diffusion term, \emph{i.e.}
\begin{equation}
  \label{eq:StuartLand}
  \frac{\partial W}{\partial \tau} =
  \lambda_2 W - g |W|^2W.
\end{equation}
If we put $W=Re^{{\rm i}\theta}$, then
\eqn(\ref{eq:StuartLand}) has the following simple solution
\begin{subeqnarray}
  \label{eq:StuartLandPolar}
  R & = & \sqrt{\frac{\sigma_2}{g'}},\\
  \theta(\tau) & = & (\omega_2 - \sigma_2\frac{g''}{g'})\tau,
\end{subeqnarray}
for the motion on the limit cycle. Expressing the solution in
terms of the real time $t$ we find that the Stuart-Landau
prediction of the oscillations of the chemical concentrations
will be described by the expression
\begin{equation}
  \label{eq:ConcHomo}
  {\bf c}(t) = \stat{\bf c} + \epsilon \frac{\sigma_2}{g'}
  \left(%
    e^{\I\omega(\mu)t}\vpa + 
    e^{-\I\omega(\mu)t}\vpb
  \right),
\end{equation}
where $\omega(\mu) = \omega_0+ (\omega_2 - \sigma_2
\frac{g''}{g'})\mu$. By plotting and comparing
\eqn(\ref{eq:ConcHomo}) together with the actual solution of
the reaction system, one normally will get an initial
estimation of whether or not the Ginzburg-Landau approach can
be justified.

\vspace{\parsep}%
As the first point of investigation, we have chosen
\chem{[\bromat] = 0.01205\;\molar} with \chem{[\hp] =
1\;\molar}, which lies on the stable side of the Benjamin-Feir
boundary (\mbox{$1+\alpha\beta = 0.76$}). Plots of the
Stuart-Landau solution and the actual solution to the
homogeneous Oregonator model are presented in
\mbox{\fig\ref{fig:SLComp}a}. The point is chosen at a
distance from the bifurcation where the amplitude of the
Stuart-Landau solution is 10\% of the value of \chem{[\cef]}
at the stationary point $\stat{\bf c}$. The results for the
reaction-diffusion equation and the CGLE on a two-dimensional
spatial domain are shown in \fig\ref{fig:StabSpiral}. One
clearly sees that the spatial wavelengths of the spirals are
in excellent agreement, \chem{4.3195\;cm} and
\chem{4.3194\;cm} respectively, determined by Fourier
transformation. The amplitudes (not shown) also agree. Note
also that the actual times it takes for the two spirals to
evolve are almost the same, which clearly justifies the time
scaling $\tau = \mu t$ predicted by the Ginzburg-Landau
theory.

\vspace{\parsep}%
In order to see what happens when the distance from the
bifurcation point is increased, we have chosen two different
values of \chem{[\bromat]} which give rise to amplitudes of
50\% and 60\% of the stationary concentration of
\chem{\cef}. The comparisons for the homogeneous systems,
\emph{i.e.} between the actual limit cycle and the
Stuart-Landau prediction are shown in \figs\ref{fig:SLComp}b
and~\ref{fig:SLComp}c. The results of the integration of the
reaction-diffusion system and the CGLE for these two working
points are presented in \figs\ref{fig:StabFifty}a
and~\ref{fig:UnsSixty}b. From \fig\ref{fig:StabFifty}a it is
evident at such large value of the bifurcation parameter
\chem{[\bromat] = 1.2950\times 10^{-2}\;\molar}, that the
spiral center in the reaction-diffusion system no longer is
localized in the center of the grid. This ``symmetry break''
is associated with the no-flux boundary conditions and we
shall ignore it in our comparisons. Even though the complex
Ginzburg-Landau equation fails in detail, it still reproduces
the behavior of the reaction-diffusion system quite well. The
spatial wavelengths are almost equal, \chem{0.772\;cm} for the
reaction-diffusion equation and \chem{0.720\;cm} for the CGLE,
and the characteristic time scales of spiral development are
essentially the same for the two systems.

\vspace{\parsep}% 
We now consider the case $\chem{[\bromat] = 1.3596 \times
10^{-2}\;\molar}$ with an amplitude 60\% of the average value
for the homogeneous Stuart-Landau solution. From
\fig\ref{fig:UnsSixty}b we clearly see that at this rather
large distance from the bifurcation point, the Ginzburg-Landau
approach breaks down. The CGLE description predicts a stable
spiral solution for the reaction-diffusion system, which
itself exhibits a totally different pattern: A small spiral
tip is formed near the boundary. After a few windings the
spiral becomes unable to maintain its structure. It becomes
unstable and breaks up into a large number of small spiral
cores distributed throughout the spatial domain of the
system. The visual appearance of the solution to the
reaction-diffusion equation does indeed look
``complex''. However, the system is not chaotic. This has been
investigated by calculating the largest Lyapunov exponent
$\lambda_{\rm max}$ which is zero. The system is thus
characterized by an oscillatory state which in this case is
dominated by a large collection of relatively small
spirals. Even though the resemblance between the
reaction-diffusion system and the CGLE seems to be lost
completely for this particular situation, one should note that
the spatial wavelength of the small reaction-diffusion spirals
still are in good agreement with the spatial wavelength found
in the single spiral wave solution to the~CGLE.

\vspace{\parsep}%
We now turn to a point on the bifurcation curve where the
Benjamin-Feir parameter \mbox{$1+\alpha\beta$} is negative,
\emph{viz.} \chem{[\bromat]_{Hopf} = 0.073\;\molar} and
\chem{[\hp]_{Hopf} = 0.348\;\molar}\ where the Benjamin-Feir
parameter equals \mbox{$1+\alpha\beta = -0.13$}. The actual
integrations of the reaction-diffusion system were carried out
at \chem{[\bromat] = 0.0817\;\molar} with \chem{[\hp] =
[\hp]_{Hopf}}. Here the Stuart-Landau amplitude is 10\% of the
average \chem{\cef} concentration, see
\fig\ref{fig:SLComp}d. The results for the reaction-diffusion
equation and the CGLE are shown in \figs\ref{fig:TurbPhase}a
(phase) and~\ref{fig:TurbPhase}b (amplitude). In both
simulations (RDE and CGLE) we see that initially a spiral wave
begins to develop from the center of the region. However, when
two spiral windings have been formed, the spiral becomes
unstable and a shock-like circular wave develops from the
spiral core. When this shock wave hits the boundaries of the
spatial domain a number of new spiral cores are generated
(these are often referred to as phaseless points). Some of
these phaseless points will repeat the scenario just
described, whereas others will be annihilated either at the
boundaries of the system or by colliding with other phaseless
points. Finally, we end up with a spatio-temporally chaotic
state which is totally dominated by phaseless points. From
\fig\ref{fig:TurbPhase}a~and~\ref{fig:TurbPhase}b we note the
striking similarity between the behavior of the
reaction-diffusion equation and the corresponding CGLE\@. The
development of the chaotic state takes place in almost the
same time in the two systems. Furthermore, we see that the
number of phaseless points in the spatial domain of the two
systems are of the same order of magnitude. The chaotic
character of the system is illustrated in
\fig\ref{fig:Lyapunov}, where the estimate of the largest
Lyapunov exponent for the CGLE system, $\lambda_{\rm max}$, is
plotted as a function of time. For $t \rightarrow \infty$ the
curve converges to a limiting value, the actual value of the
Lyapunov exponent. As is evident from \fig\ref{fig:Lyapunov},
the Lyapunov exponent is positive corresponding to chaotic
dynamics.

\vspace{\parsep}%
The characteristic parameters used for the numerical
integrations discussed in this section are summarized in
table~\ref{tab:BifDist}.

\section{Computational details}
The computation of the Hopf bifurcation diagram and the limit
cycle solutions to the homogeneous Oregonator model were done
with the bifurcation analysis package \emph{CONT}
\citep{Cont}. All integrations of the reaction-diffusion
systems were performed on a CRAY92-computer using an explicit
fourth order Runge-Kutta algorithm~\citep{NumRecC}. The
integration of the CGLE were done on a HPUX-workstation using
the implicit Adams method from ODEPACK~\citep{Lsode}. All
integrations were carried out with no-flux boundary conditions
and initial conditions chosen as described in \cite[p.\
106]{Kuramoto}. The Lyapunov exponent was computed by using a
numerical approach described by Marek and
Schreiber~[\citeyear{Marek91}].

\section{Discussion}
We consider two ways of using the complex Ginzburg-Landau
equation for chemical reaction diffusion systems. One is to
simplify calculations with models.  Another one is to base
realistic modelling of actual reaction diffusion systems on
the CGLE with parameters determined directly by experiments
performed on homogeneous systems [Hynne \& S{\o}rensen, 1993].

\vspace{\parsep}%
The numerical simplification of the reaction-diffusion models
by the CGLE is obtained in two ways. First, the CGLE
effectively reduces the dimension of the state space to two,
which saves computation time, particularly for
high-dimensional models.  For example, calculation of waves
for the big model of Gy\"orgyi~\emph{et
al.}~[\citeyear{BigBZModel}] with its 26-dimensional state
space, is no more difficult numerically than for any other
model --- once the Ginzburg-Landau parameters have been
calculated. (Here the main problem is to obtain all the
relevant diffusion coefficients). Second and most importantly,
the use of the CGLE solves the problem (with the RDE) of the
critical slowing down of pattern development that occurs near
a bifurcation. The CGLE works with amplitudes, and the major
effect of the time oscillations is accounted for analytically
(in the transformation to the amplitude equation) rather than
numerically. In fact, the critical slowing down has been
completely scaled away in the CGLE. The importance of this
feature can be appreciated by the significant reductions (by
factors up to 1000) of the number of steps necessary for the
integration, see \tabl\ref{tab:BifDist}.  In practice this
means that the CGLE-calculation can be made with a work
station whereas a supercomputer may be needed for integrating
the RDE. In addition, the CGLE provides the solution in the
form of a time and space dependent complex amplitude. This
form is very convenient for understanding solutions, and it
can easily be transformed back to actual time and space
dependent concentrations of the (multitude of) species
participating in the reactions.

\vspace{\parsep}%
The other way of using the CGLE for chemical systems is to set
up the Ginzburg-Landau equation \emph{directly} from
experimental measurements on the corresponding {\em
homogeneous} reaction system using independently measured
diffusion coefficients. It has been shown previously
\citep{PgsCgle} how the parameters can be obtained from
quenching experiments \citep{CstrFig,GenQuench,Vladana,Nagy}.
Thus, it is possible to model waves and chaos in real spatial
systems without knowing the detailed mechanism of the chemical
reactions.

\vspace{\parsep}%
The previous discussion shows that the Ginzburg-Landau
approach to chemical reaction-diffusion systems is potentially
very important. However, the use of the CGLE depends on the
validity of the approximation for any particular system and
operating point.

\vspace{\parsep}%
To throw some light on the question of the validity of the
complex Ginzburg-Landau approximation we have compared
solutions to a model reaction-diffusion system with their
CGLE-approximations. For a chemical reaction diffusion system
based on the Oregonator model of the Belousov-Zhabotinsky
reaction, we find that there exist substantial regions for
which the CGLE provides an excellent description of the
dynamical behavior. This conclusion applies to parameter
regions where the solutions show spatio-temporal chaos as well
as regions with stable spiral waves. In the latter case, we
demonstrate how the CGLE approximation eventually fails beyond
a certain limit.

\vspace{\parsep}%
Experimentally, it is difficult to work too close to a Hopf
bifurcation because small amplitude waves are difficult to
detect and because of slow pattern evolution (critical slowing
down). We conclude from the present study that one can
reasonably expect a Ginzburg-Landau description to remain
applicable to a system at experimentally feasible distances
from the bifurcation point, provided the properties of the
Oregonator model can be regarded as representative of the
investigated system.

\vspace{\parsep}%
One condition for using the conclusions of the present study
is that the real parts of the eigenvalues of the Jacobian
matrix (other than the bifurcating pair) are sufficiently
negative as in the Oregonator model considered here so that
excitations out of the plane of oscillations decay fast. We
are currently studying what happens when one such "transient
mode" is slow, and the present work serves as a preliminary to
that study as well as a prerequisite to using the CGLE with
experiments and models.

\vspace{\parsep}%
To conclude, waves and spatio-temporal chaos in real chemical
systems can be described by a complex Ginz\-burg-Lan\-dau
equation with parameters $\alpha$ and $\beta$ determined by
quenching experiments performed on the specific reaction at
the specific point of operation used. The paper has
demonstrated that, typically, the Ginzburg-Landau description
will apply in a sufficiently wide region surrounding the
bifurcation set to make the description useful in practice.

%%%%%%%%%%%%%%%%%%%%%%%%%%%%%%%%%%%%%%%%%%%%%%%%%%%%%%%%%%%
%% Here Come the Warm References!
%%%%%%%%%%%%%%%%%%%%%%%%%%%%%%%%%%%%%%%%%%%%%%%%%%%%%%%%%%%
\newpage
\bibliographystyle{bifchaos}
\bibliography{dynamic}

%%%%%%%%%%%%%%%%%%%%%%%%%%%%%%%%%%%%%%%%%%%%%%%%%%%%%%%%%%%
%% List of table and figure captions.
%%%%%%%%%%%%%%%%%%%%%%%%%%%%%%%%%%%%%%%%%%%%%%%%%%%%%%%%%%%
\newpage
\Large\textbf{Captions of tables and figures}
\normalsize

\vspace{2.5mm}%
\textbf{Table~\ref{tab:RateVals}}: Rate constants
$k_1,\ldots,k_5$, stoichiometric factor $f$ and diffusion
constants $D_X$, $D_Y$ and $D_Z$ used in the numerical
integration of the Oregonator model,
\eqn(\ref{eq:OrgDiff}). The values of $k_1,\ldots,k_4$ are
from Field and F\"{o}rsterling~[\protect\citeyear{FFrate}],
$k_5$ and $f$ from Nielsen \emph{et al.}
[\protect\citeyear{HopfQuench}] and diffusion constants from
Hynne \& Sørensen~[\citeyear{PgsCgle}]. (\chem{\molar =
mol\;dm^{-3}}).

\vspace{4.0mm}%
\textbf{Table~\ref{tab:BifVals}}: Parameter values of the
complex Ginzburg-Landau equation at the two points of the
$(1+\alpha\beta)$-curve plotted in \fig\ref{fig:BFPlot}
corresponding to the values of the two-parameter point
$(\chem{[\bromat]_{\rm Hopf}},\chem{[\hp]_{\rm Hopf}})$. The
values of \chem{[\bromat]} actually used in the numerical
integration of the reaction-diffusion system are given in
\tabl\ref{tab:BifDist}. Note that these values are off the
bifurcation curve. The number $\omega_0$ is the frequency of
the sinusoidal oscillations exactly at the Hopf bifurcation
point, whereas $\sigma_2$ and $\omega_2$ are the derivatives
of the real and imaginary parts of the eigenvalue $\lambda$
with respect to the bifurcation parameter (\emph{i.e.}
\chem{[\bromat]} for this particular example). The parameters
$g$ and $d$ are the coefficients of the nonlinear term and the
diffusion term in the complex Ginzburg-Landau equation, from
which the dimensionless parameters $\alpha$ and $\beta$ are
calculated (see
\eqns(\ref{eq:GinzLand})~and~(\ref{eq:TransCGLE})). The
Benjamin-Feir parameter $1+\alpha\beta$ for the two Hopf
bifurcation points have been selected in order to study two
different situations: At the first point ($1+\alpha\beta>0$)
we find that the reaction-diffusion system to admits of stable
spiral wave solutions, whereas for the second point
($1+\alpha\beta<0$) we find chaotic behavior.

\vspace{4.0mm}%
\textbf{Table~\ref{tab:BifDist}}: Integration parameters for
the reaction-diffusion equation and the corresponding complex
Ginzburg-Landau equation for different amplitudes relative to
the stationary $\cef$ concentration and for the two selected
points on the Hopf bifurcation curve
(\fig\ref{fig:BifDiagram}). The table shows the final
integration time, $t_{\rm end}$, and size, $(r_x,r_y)$, of the
2-dimensional domain for the specific bifurcation point
distances of \chem{[\bromat]} used in the integration of the
reaction-diffusion system. $\tau_{\rm end}$ and $(s_x,s_y)$
indicate the corresponding scaled values used in the
integration of the CGLE. The size of the grid is also shown
together with a comparison between the steps used by the
solver to reach $t_{\rm end}$ for the reaction-diffusion
system and the associated CGLE (equal grid sizes was chosen
for the integration of a specific reaction-diffusion system
and the corresponding CGLE). 

\vspace{4.0mm}%
\textbf{Figure~\ref{fig:BifDiagram}}: Bifurcation diagram
showing the locations of Hopf bifurcations in the Oregonator
model in the plane of the two parameters \chem{[\bromat]} and
\chem{[\hp]}. At a specific point
\chem{[\bromat]=0.269\;\molar} and \chem{[\hp]=0.176\;\molar},
the Hopf bifurcation changes stability from supercritical to
subcritical corresponding to the solid and dashed curves
respectively. For each point on the solid curve, a unique
Ginzburg-Landau equation is defined, since the Ginzburg-Landau
parameters $\omega_2$, $\sigma_2$, $g$ and $d$ can be
calculated from
\eqns(\ref{eq:Solv1})--(\ref{eq:SolvCond}). The comparisons
between the reaction-diffusion system and the complex
Ginzburg-Landau equation have been carried out at the two
points indicated on the curve.

\vspace{4.0mm}%
\textbf{Figure~\ref{fig:AlphaPlot}}: (a): The dimensionless
diffusion parameter $\alpha$ along the supercritical branch in
\fig\ref{fig:BifDiagram}, plotted as a function of the
parameter \chem{[\bromat]}. (b): Variation of the
dimensionless parameter $\beta$ associated with the nonlinear
term in the complex Ginzburg-Landau equation along the
supercritical branch in \fig\ref{fig:BifDiagram} plotted as a
function of the parameter \chem{[\bromat]}. 

\vspace{4.0mm}%
\textbf{Figure~\ref{fig:BFPlot}}: The figure shows the
variation of the Benjamin-Feir stability parameter
$1+\alpha\beta$ along the curve of Hopf bifurcation points in
the Oregonator model with \chem{[\bromat]} (a) and
\chem{[\hp]} (b) as parameter. From (a) and (b) we see that
the Oregonator reaction-diffusion system crosses the
Benjamin-Feir instability border at the point \chem{[\bromat]
= 0.073\;\molar} and \chem{[\hp] = 0.348\;\molar} where
$1+\alpha\beta$ changes sign. As reference for further
numerical investigations we have chosen two points indicated
on figure (a) and (b) with a filled and an open circle: One
point showing stable spiral wave solutions and one point
exhibiting spatio-temporally chaotic behavior.

\vspace{4.0mm}%
\textbf{Figure~\ref{fig:SLComp}}: Comparisons of the actual
periodic solutions to the homogeneous Oregonator model (solid
lines) and its approximation by the Stuart-Landau equation
(dashed lines) at various distances from the bifurcation
point. (a)-(c): Periodic solutions with amplitudes of 10\%,
50\% and 60\% of the stationary concentration of \chem{\cef}
at the Hopf bifurcation point represented in
\fig\ref{fig:BifDiagram} with a filled circle.  (d):
Oscillations with an amplitude of 10\% of the stationary
concentration of \chem{\cef} associated with the Hopf
bifurcation point represented with an open circle in
\fig\ref{fig:BifDiagram}.

\vspace{4.0mm}%
\textbf{Figure~\ref{fig:StabSpiral}}: The phase $\phi$ of the
oscillations for a stable spiral solution to the Oregonator
reaction-diffusion system (RDE) and the associated complex
Ginzburg-Landau equation (CGLE). The parameters used in the
integrations correspond to the point marked with a filled
circle in \figs\ref{fig:BifDiagram}~and~\ref{fig:BFPlot}. In
this case there is a very good agreement between the time
scale under which the two patterns develop as well as the
spatial wavelength of the two spirals. The numbers below each
of the six snapshots of the RDE shows the time elapsed for
that particular state in the RDE in units of
\chem{10^5\;s}. The corresponding snapshots for the CGLE are
made at corresponding scaled times. The reaction-diffusion
system has been solved for the parameter value
\chem{[\bromat]=1.2051 \times 10^{-2}\;\molar} corresponding
to an amplitude of 10\% of the stationary concentration of
\chem{\cef} for the homogeneous Oregonator system.

\vspace{4.0mm}%
\textbf{Figure~\ref{fig:StabFifty}}: Deviations between the
reaction-diffusion system and the CGLE description become more
evident when a larger distance from the bifurcation point is
chosen. The figure compares the two systems at two finite
distances corresponding to an amplitude of 50\% (a) and 60\%
(b) of the stationary $\cef$ concentration. Due to nonlinear
effects which become dominating as the distance increases the
spiral moves away from the center of the domain and freezes at
a point close to the boundary of the domain. In (a) the
general characteristics (spatial wavelength and time of
evolution) of the spirals in the two systems are still in good
agreement. When the distance is increased further (b) the
resemblance between the reaction-diffusion system and the CGLE
is completely lost. The final state of the reaction-diffusion
system is characterized by a number of small spirals
distributed throughout the domain of the system. However, the
spatial wavelength of these spirals are still described well
by the CGLE.

\vspace{4.0mm}%
\textbf{Figure~\ref{fig:TurbPhase}}: Development of a chaotic
state in the Oregonator reaction-diffusion system (RDE) and
the associated complex Ginzburg-Landau equation (CGLE)
represented by the phase $\phi$ (a) and the amplitude (b) of
the oscillations of \chem{[\cef]}. The characteristic
parameter values used in the RDE and the CGLE correspond to
the parameter point marked with an open circle in
\figs\ref{fig:BifDiagram}~and~\ref{fig:BFPlot} where the
Benjamin-Feir parameter satisfies
$1+\alpha\beta<0$. Indications below the twelve snapshots of
the RDE shows the evolution time for that particular state in
the RDE as well as for the equivalent state in the CGLE
measured in units of \chem{10^5\;s}. Initially a spiral is
formed at the center of the grid, but after approximately two
windings have been, formed this structure breaks down being
replaced by a set of phaseless points. (A phaseless point
appears as a very localized dark spot on the
\chem{[\cef]}-plot). The phaseless points will either split
into other phaseless points or be annihilated at the boundary
of the spatial domain. This scenario is then repeated,
resulting in a very complicated interaction between the
phaseless points. Finally a spatio-temporally chaotic state is
reached in both the RDE and the CGLE\@. In the plots of the
phase and the amplitude of \chem{[\cef]} we see a striking
similarity between the behavior exhibited by the RDE and the
CGLE\@. The two patterns develop in an almost synchronized
manner. Even when the spiral is fully destroyed (\chem{{\it t}
= 2.5 \times 10^5\;s}) the positions of the phaseless points
are still very similar in the corresponding RDE and CGLE
states. The synchronization is however lost at the final
chaotic state, due to the fact that this chaotic state is
characterized by a critical dependence on the initial
conditions. The reaction-diffusion system has been solved for
the parameter value \chem{[\bromat]=8.1935 \times
10^{-2}\;\molar} corresponding to an amplitude of 10\% of the
stationary $\cef$ concentration.

\vspace{4.0mm}%
\textbf{Figure~\ref{fig:Lyapunov}}: Largest Lyapunov exponent,
$\lambda_{\rm max}$, calculated for the complex
Ginzburg-Landau equation for the state presented in
\fig\ref{fig:TurbPhase} (shown in
\figs\ref{fig:BifDiagram}~and~\ref{fig:BFPlot} with an open
circle). The value of $\lambda_{\rm max}$ is plotted as a
function of time for a point where $1+\alpha\beta < 0$ giving
rise to spatio-temporal chaos and a positive Lyapunov
exponent. In this sense, the state of the system can be
characterized as chaotic.

\newpage
\pagestyle{empty}
%%%%%%%%%%%%%%%%%%%%%%%%%%%%%%%%%%%%%%%%%%%%%%%%%%%%%%%%%%%
%% The Oregonator constants. --- Shown in tabular form
%%%%%%%%%%%%%%%%%%%%%%%%%%%%%%%%%%%%%%%%%%%%%%%%%%%%%%%%%%%
\begin{table}[htbp]
  \begin{center}
    \leavevmode
    \begin{tabular}{||l|l||}\hline
      \emph{Constant} & \emph{Value} \\ \hline\hline
      $k_1/\chem{\molar^{-3}s^{-1}}$  & $2.0$ \\ \hline
      $k_2/\chem{\molar^{-2}s^{-1}}$  & $3.0 \times 10^{6}$ \\ \hline
      $k_3/\chem{\molar^{-1}s^{-1}}$  & $3.0 \times 10^{3}$ \\ \hline
      $k_4/\chem{\molar^{-2}s^{-1}}$  & $42.0$              \\ \hline
      $k_5/\chem{s^{-1}}$        & $0.167$             \\ \hline
      $f$                        & $0.7905$            \\ \hline\hline
      $D_X/\chem{cm^2s^{-1}}$ & $1.0\times 10^{-5}$    \\ \hline
      $D_Y/\chem{cm^2s^{-1}}$ & $1.6\times 10^{-5}$    \\ \hline
      $D_Z/\chem{cm^2s^{-1}}$ & $0.6\times 10^{-5}$    \\ \hline
    \end{tabular}
  \end{center}
  \caption{}
  \label{tab:RateVals}
\end{table}

%%%%%%%%%%%%%%%%%%%%%%%%%%%%%%%%%%%%%%%%%%%%%%%%%%%%%%%%%%%
%% Bifurcation Data. --- Shown in tabular form
%%%%%%%%%%%%%%%%%%%%%%%%%%%%%%%%%%%%%%%%%%%%%%%%%%%%%%%%%%%
\newcommand{\brcs}{\chem{1.2014 \times 10^{-2}}}
\newcommand{\brct}{\chem{8.1765 \times 10^{-2}}}
\newcommand{\hpcs}{\chem{1.0000}}
\newcommand{\hpct}{\chem{3.2721 \times 10^{-1}}}
\newcommand{\brs} {\chem{1.2051 \times 10^{-2}}}
\newcommand{\brt} {\chem{8.1935 \times 10^{-2}}}
\newcommand{\omos}{\chem{0.1085}}
\newcommand{\omot}{\chem{0.1200}}

\newcommand{\sis} {\chem{3.4468}}
\newcommand{\sit} {\chem{0.6000}}
\newcommand{\omts}{\chem{5.3686}}
\newcommand{\omtt}{\chem{0.8196}}

\newcommand{\gs}{\chem{4.08 + {\rm i} 3.62}}
\newcommand{\gt}{\chem{6.54 + {\rm i} 2.02}}
\newcommand{\ds}{\chem{1.01 - {\rm i} 0.28}}
\newcommand{\dt}{\chem{1.03 - {\rm i} 0.38}}
\newcommand{\als}{\chem{-0.2712}}
\newcommand{\alt}{\chem{-0.3651}}
\newcommand{\bes}{\chem{0.8846}}
\newcommand{\bet}{\chem{3.1020}}

\newcommand{\bfs}{\chem{0.76010}}
\newcommand{\bft}{\chem{-0.13237}}

\begin{table}[htbp]
  \small
  \begin{center}
    \leavevmode
    \begin{tabular}{||l|l|l||}\hline
      \emph{Parameter} & \emph{Spiral}  & \emph{Turbulence}\\\hline\hline
      \chem{[\bromat]_{Hopf}/\molar}    & \brcs & \brct \\ \hline
      \chem{[\hp]_{Hopf}/\molar}        & \hpcs & \hpct \\ \hline
      $\omega_0/\chem{s^{-1}}$          & \omos & \omot \\ \hline
      $\sigma_2/\chem{s^{-1}}$          & \sis  & \sit  \\ \hline
      $\omega_2/\chem{s^{-1}}$          & \omts & \omtt \\ \hline
      $g/\chem{10^{13}\molar^{-2}s^{-1}}$    & \gs   & \gt   \\ \hline
      $d/\chem{10^{-5}cm^2s^{-1}}$      & \ds   & \dt   \\ \hline
      $\alpha$   & \als  & \alt     \\ \hline
      $\beta$    & \bes  & \bet     \\ \hline
      $1+\alpha\beta$ & \bfs & \bft \\ \hline
    \end{tabular}
  \end{center}
  \caption{}
  \label{tab:BifVals}
  \normalsize
\end{table}

\newpage
%%%%%%%%%%%%%%%%%%%%%%%%%%%%%%%%%%%%%%%%%%%%%%%%%%%%%%%%%%%
%% Table of Distances from Bifurcation points.
%%%%%%%%%%%%%%%%%%%%%%%%%%%%%%%%%%%%%%%%%%%%%%%%%%%%%%%%%%%
\newcommand{\brsa}{\chem{1.2051 \times 10^{-2}}}
\newcommand{\brsb}{\chem{1.2950 \times 10^{-2}}}
\newcommand{\brsc}{\chem{1.3596 \times 10^{-2}}}
\newcommand{\brta}{\chem{8.1765 \times 10^{-2}}}

\newcommand{\musa}{\chem{3.7456 \times 10^{-5}}}
\newcommand{\musb}{\chem{9.3642 \times 10^{-4}}}
\newcommand{\musc}{\chem{1.3484 \times 10^{-3}}}
\newcommand{\muta}{\chem{1.7082 \times 10^{-4}}}

\newcommand{\trsa}{\chem{6.50}}
\newcommand{\trsb}{\chem{0.96}}
\newcommand{\trsc}{\chem{0.44}}
\newcommand{\trta}{\chem{3.30}}

\newcommand{\ttsa}{\chem{24.3}}
\newcommand{\ttsb}{\chem{89.9}}
\newcommand{\ttsc}{\chem{59.3}}
\newcommand{\ttta}{\chem{56.4}}

\newcommand{\rrsa}{\chem{(30.0,30.0)}}
\newcommand{\rrsb}{\chem{(10.0,10.0)}}
\newcommand{\rrsc}{\chem{(5.0,5.0)}}
\newcommand{\rrta}{\chem{(20.0,20.0)}}

\newcommand{\rssa}{\chem{(0.184,0.184)}}
\newcommand{\rssb}{\chem{(0.306,0.306)}}
\newcommand{\rssc}{\chem{(0.198,0.198)}}
\newcommand{\rsta}{\chem{(0.261,0.261)}}

\newcommand{\gssa}{\texttt{128x128}}
\newcommand{\gssb}{\texttt{256x256}}
\newcommand{\gssc}{\texttt{256x256}}
\newcommand{\gsta}{\texttt{128x128}}

\newcommand{\sssa}{\chem{6.5 \times 10^5}/\chem{649}}
\newcommand{\sssb}{\chem{1.9 \times 10^5}/\chem{9600}}
\newcommand{\sssc}{\chem{8.8 \times 10^4}/\chem{3747}}
\newcommand{\ssta}{\chem{3.3 \times 10^5}/\chem{3300}}

\begin{table}[htbp]
  \small
  \begin{center}
    \leavevmode
    \begin{tabular}{||l|l|l|l||l||}\cline{2-5}
      \multicolumn{1}{c}{} &
      \multicolumn{3}{|c||}{\emph{Spiral}} &
      \multicolumn{1}{c||}{\emph{Turbulence}}\\\hline
      \emph{amplitude deviation}  & 10\%  & 50\% & 60\% & 10\% \\\hline\hline
      \chem{[\bromat]/\molar}     &\brsa  &\brsb &\brsc &\brta \\\hline
      \chem{(\mu-\mu_0)/\molar}   &\musa  &\musb &\musc &\muta \\\hline
      \chem{t_{end}/10^{5}s}      &\trsa  &\trsb &\trsc &\trta \\\hline
      \chem{\tau_{end}}           &\ttsa  &\ttsb &\ttsc &\ttta \\\hline
      \chem{(r_x,r_y)/cm}         &\rrsa  &\rrsb &\rrsc &\rrta \\\hline
      \chem{(s_x,s_y)}            &\rssa  &\rssb &\rssc &\rsta \\\hline
      grid                        &\gssa  &\gssb &\gssc &\gsta \\\hline
      steps - RDE/CGLE            &\sssa  &\sssb &\sssc &\ssta \\\hline
    \end{tabular}
  \end{center}
  \caption{}
  \label{tab:BifDist}
  \normalsize
\end{table}

%%%%%%%%%%%%%%%%%%%%%%%%%%%%%%%%%%%%%%%%%%%%%%%%%%%%%%%%%%%
%% Bifurcation diagram.
%%%%%%%%%%%%%%%%%%%%%%%%%%%%%%%%%%%%%%%%%%%%%%%%%%%%%%%%%%%
\begin{figure}[htbp]
  \begin{center}
    \leavevmode
    \vspace{5.0cm}
    \normalsize
  \end{center}
  \caption{}
  \label{fig:BifDiagram}
\end{figure}

%%%%%%%%%%%%%%%%%%%%%%%%%%%%%%%%%%%%%%%%%%%%%%%%%%%%%%%%%%%
%% Alpha and Beta plot.
%%%%%%%%%%%%%%%%%%%%%%%%%%%%%%%%%%%%%%%%%%%%%%%%%%%%%%%%%%%
\begin{figure}[htbp]
  \begin{center}
    \leavevmode
    \vspace{5.0cm}
    \vspace{1.5cm}
    \normalsize
  \end{center}
  \caption{}
  \label{fig:AlphaPlot}
\end{figure}

%%%%%%%%%%%%%%%%%%%%%%%%%%%%%%%%%%%%%%%%%%%%%%%%%%%%%%%%%%%
%% Benjamin-Feir plot.
%%%%%%%%%%%%%%%%%%%%%%%%%%%%%%%%%%%%%%%%%%%%%%%%%%%%%%%%%%%
\begin{figure}[htbp]
  \begin{center}
    \leavevmode
    \vspace{5.0cm}
    \vspace{1.0cm}
    \normalsize
  \end{center}
  \caption{}
  \label{fig:BFPlot}
\end{figure}

%%%%%%%%%%%%%%%%%%%%%%%%%%%%%%%%%%%%%%%%%%%%%%%%%%%%%%%%%%%
%% Comparison between the SLE and the Oregonator.
%%%%%%%%%%%%%%%%%%%%%%%%%%%%%%%%%%%%%%%%%%%%%%%%%%%%%%%%%%%
\begin{figure}[htbp]
  \begin{center}
    \leavevmode
    \vspace{5.0cm}
    \normalsize
  \end{center}
  \caption{}
  \label{fig:SLComp}
\end{figure}

%%%%%%%%%%%%%%%%%%%%%%%%%%%%%%%%%%%%%%%%%%%%%%%%%%%%%%%%%%%
%% Stabil spiral: CGLE (10%)
%%%%%%%%%%%%%%%%%%%%%%%%%%%%%%%%%%%%%%%%%%%%%%%%%%%%%%%%%%%
\begin{figure}[t]
  \begin{center}
    \leavevmode
    \small
    \vspace{5.0cm}
  \end{center}
  \normalsize
  \caption{}
  \label{fig:StabSpiral}
\end{figure}
\clearpage

%%%%%%%%%%%%%%%%%%%%%%%%%%%%%%%%%%%%%%%%%%%%%%%%%%%%%%%%%%%
%% Stabil spiral: CGLE (50%)
%%%%%%%%%%%%%%%%%%%%%%%%%%%%%%%%%%%%%%%%%%%%%%%%%%%%%%%%%%%
\begin{figure}[t]
  \begin{center}
    \leavevmode
    \small
    \vspace{5.0cm}
  \end{center}
  \normalsize
  \caption{}
  \label{fig:StabFifty}
\end{figure}
\clearpage

\addtocounter{figure}{-1}
%%%%%%%%%%%%%%%%%%%%%%%%%%%%%%%%%%%%%%%%%%%%%%%%%%%%%%%%%%%
%% Stabil spiral: CGLE (60%)
%%%%%%%%%%%%%%%%%%%%%%%%%%%%%%%%%%%%%%%%%%%%%%%%%%%%%%%%%%%
\begin{figure}[t]
  \begin{center}
    \leavevmode
    \small
    \vspace{5.0cm}
  \end{center}
  \normalsize
  \caption{}
  \label{fig:UnsSixty}
\end{figure}
\clearpage

%%%%%%%%%%%%%%%%%%%%%%%%%%%%%%%%%%%%%%%%%%%%%%%%%%%%%%%%%%%
%% Turbulence: Phase (2dim).
%%%%%%%%%%%%%%%%%%%%%%%%%%%%%%%%%%%%%%%%%%%%%%%%%%%%%%%%%%%
\begin{figure}[t]
  \begin{center}
    \leavevmode 
    \small 
    \vspace{5.0cm}
  \end{center}
  \normalsize
  \caption{}
  \label{fig:TurbPhase}
\end{figure}
\clearpage

\addtocounter{figure}{-1}
%%%%%%%%%%%%%%%%%%%%%%%%%%%%%%%%%%%%%%%%%%%%%%%%%%%%%%%%%%%
%% Turbulence: Amplitude (2dim).
%%%%%%%%%%%%%%%%%%%%%%%%%%%%%%%%%%%%%%%%%%%%%%%%%%%%%%%%%%%
\begin{figure}[t]
  \begin{center}
    \leavevmode
    \small
    \vspace{5.0cm}
  \end{center}
  \normalsize
  \caption{}
  \label{fig:TurbAmpl}
\end{figure}
\clearpage

%%%%%%%%%%%%%%%%%%%%%%%%%%%%%%%%%%%%%%%%%%%%%%%%%%%%%%%%%%%
%% Largest Lyapunov exponent.
%%%%%%%%%%%%%%%%%%%%%%%%%%%%%%%%%%%%%%%%%%%%%%%%%%%%%%%%%%%
\begin{figure}[t]
  \begin{center}
    \leavevmode
    \vspace{5.0cm}
  \end{center}
  \caption{}
  \label{fig:Lyapunov}
\end{figure}

\end{document}